\documentclass[fleqn,10pt]{wlscirep}
\usepackage{subfigure}
\usepackage{amsmath}
\usepackage{mathtools}
\usepackage{soul}

\newcommand{\fref}[1]{Fig.\ref{#1}}
\newcommand{\rb}{$^{87}$Rb}
\newcommand{\Ehfs}{E_{\mathrm{hfs}}}

\title{{$\Lambda$}-enhanced grey molasses on the $D_2$ transition of Rubidium-87 atoms  }

\author[1,2]{Sara Rosi}
\author[1,2]{Alessia Burchianti}
\author[1]{Stefano Conclave}
\author[1,2,+]{Devang S. Naik}
\author[1,2]{Giacomo Roati}
\author[1,2]{Chiara Fort}
\author[1,2,*]{Francesco Minardi}
\affil[1]{LENS European Laboratory for Non-Linear Spectroscopy, and Dipartimento di Fisica e Astronomia, Università di Firenze, Sesto Fiorentino, 50019, Italy}
\affil[2]{Istituto Nazionale di Ottica, INO-CNR, Sesto Fiorentino, 50019, Italy}

\affil[*]{francesco.minardi@ino.it}

\affil[+]{Current address: LP2N, Laboratoire Photonique, Num\'erique et
  Nanosciences,\\ Universit\'e Bordeaux-IOGS-CNRS, F–33400 Talence, France.}


\begin{abstract} 
Laser cooling based on dark states, i.e. states decoupled from light, has proven to be effective to increase the phase-space density of cold trapped atoms. Dark-states cooling requires open atomic transitions, in contrast to the ordinary laser cooling used for example in magneto-optical traps (MOTs), which operate on closed atomic transitions. For alkali atoms, dark-states cooling is therefore commonly operated on the $D_1$ transition $n S_{1/2}\rightarrow n P_{1/2}$. We show that, for \rb, thanks to the large hyperfine structure separations the use of this transition is not strictly necessary and that ``quasi-dark state'' cooling is efficient also on the $D_2$ line, $5 S_{1/2}\rightarrow 5 P_{3/2}$. We report temperatures as low as $(4.0\pm 0.3)\,\mu$K
and an increase of almost an order of magnitude in the phase space density with respect to ordinary laser sub-Doppler cooling.
\end{abstract}
\begin{document}

\flushbottom
\maketitle
\thispagestyle{empty}

\section*{Introduction}
 
Providing several orders of magnitude of gain in phase-space density from a room-temperature atomic
vapour, laser cooling is essential to almost all quantum gases experiments. Sub-Doppler cooling, i.e. cooling below the limit temperature of
two-level atoms, relies on a combination of ac-Stark shifts and optical pumping
among Zeeman sublevels \cite{dalibard_laser_1989}.  While intense research on the cooling mechanisms has
taken place in the '80s and early '90s, the advent of Bose-Einstein condensation in dilute alkali atoms 
diverted much of the interest of atomic physicists and laser cooling gradually
turned into a tool, ordinarily used and partially understood. Recently, interest in the 
fundamentals of laser cooling  has been revived by the demonstration of effective optical schemes for high-resolution imaging of individual atoms \cite{bakr_quantum_2009,sherson_single-atom-resolved_2010,lester_raman_2014,haller_single-atom_2015,cheuk_quantum-gas_2015,parsons_site-resolved_2015,edge_imaging_2015}, as well as direct laser cooling processes towards quantum degeneracy without any evaporative cooling stage \cite{stellmer_laser_2013,hu_creation_2017}.  
Many of these techniques employ open transitions \cite{haller_single-atom_2015,grier_lambda-enhanced_2013}. In fact, cooling on open
transitions optically pumps atoms in Zeeman dark states thereby reducing the number of
spontaneously emitted photons. Such photons impart a randomly directed recoil to the atoms, that limits the lowest
attainable temperature, and generate an effective interatomic repulsion, that limits the
highest attainable density \cite{sesko_behavior_1991}. In addition, atoms in excited states cause light-induced losses due to fine-structure changing collisions and radiative escape \cite{weiner_experiments_1999}. Such effects are detrimental to ultracold atoms experiments, for example they limit the reachable phase space density ($PSD$) in magneto-optical traps (MOTs), hindering the subsequent transfer of the atomic sample into optical or magnetic traps. 

The existence of ``dark'' states in open transitions $F \rightarrow F'$ between
ground $F$ and excited $F' (\leq F)$ hyperfine manifolds entails the possibility
for atoms to decouple from the laser light. Dark states are linear
superpositions of $|F, m_F \rangle$ Zeeman sublevels, that depend on the local
polarisation of the laser fields. Atoms moving at sufficiently low velocities
remain adiabatically dark, as the linear superposition adjusts to the slowly
varying polarisation. Instead, faster atoms undergo diabatic transitions towards
``bright'' states. With blue-detuned light, atoms are more likely to decelerate
when in bright states, thus they progressively accumulate
near the zero-velocity dark state \cite{weidemuller_novel_1994,grynberg_proposal_1994}. Such cooling has been investigated
since the late '90s and is commonly referred to as ``grey molasses'' since
it involves states neither bright nor completely dark \cite{boiron_three-dimensional_1995,esslinger_purely_1996}. Recently a twist has
been added to the picture \cite{grier_lambda-enhanced_2013}: with an additional laser frequency tuned on the
repumper transition in $\Lambda$-configuration, the dark
states become a superposition involving both $F-1$ and $F$ hyperfine levels,
dominated by the $F-1$ level when, as usually the case, the intensity of the light on the cooler transition
$F\rightarrow F'$ is much larger than the one of the repumper
transition $F-1 \rightarrow F'$ \cite{arimondo_coherent_1996}. It has been shown with $^6$Li \cite{sievers_simultaneous_2015,burchianti_efficient_2014}, $^7$Li \cite{grier_lambda-enhanced_2013}, $^{39}$K \cite{salomon_gray_2013,nath_quantum-interference-enhanced_2013}, $^{40}$K \cite{fernandes_sub-doppler_2012,sievers_simultaneous_2015}, $^{41}$K \cite{chen_production_2016}
and $^{23}$Na \cite{colzi_grey_2016} that $\Lambda$-enhanced grey cooling leads to substantial
advantages in terms of lower temperature and higher phase-space density.

With some alkali atoms, such as Li and K, the hyperfine energy separations in
the upper level of the $D_2$ line ($n S_{1/2} \rightarrow n P_{3/2}$) are of the
same order as the natural linewidths, thus the closed transitions $F\rightarrow F+1$ is
hardly isolated from the open transitions. For this reason, for such atoms grey
molasses are tipically implemented on the $D_1$ line, with the recent exception of Ref. \cite{bruce_sub-doppler_2017}. However,
other atoms, such as Rb and Cs, feature $n P_{3/2}$ hyperfine separations much
larger than the natural linewidths. For these atoms it is worth exploring grey molasses
on the $D_2$ transition that is used for the MOT, with the distinct advantage of avoiding the additional laser
source needed to implement grey molasses on the $D_1$ line.

In this work, we characterise sub-Doppler cooling in \rb\ with blue-detuned light in a wide range of frequencies blue-detuned with respect to the $F=2\rightarrow F'=2$ open transition (grey molasses).  We show that in our experiment the grey molasses
reduces the final temperatures by a factor of 4 with respect to the bright
molasses, with a minimum observed temperature of $(4.0\pm0.3)\,\mu$K in a sample of
$\sim 10^8$ atoms. In addition, $PSD$ is increased by an order of
magnitude. These results represent an important advancement for the production of quantum degenerate
gases, where laser cooling is most often followed by evaporative cooling, which greatly benefits from beginning at high $PSD$. Furthermore, the method implemented here can be useful in all experiments using \rb\ as ``coolant'' species to realize ultracold atomic mixtures by sympathetic cooling \cite{modugno_bose-einstein_2001,papp_tunable_2008,wang_double_2016}.

\section*{Results}

In this section, after a brief description of the experiment, we report the characterisation of the grey
molasses cooling procedure. For more details about the experimental procedure we refer the reader to the Methods. 

We load $N_{MOT}=3 \times 10^8$ atoms in a MOT at $100\,\mu$K from a cold atomic beam in typically 7 s. After the MOT loading, we can increase the $PSD$ of the sample adding a bright molasses stage, which gives as best result $1.5\times 10^{8}$ atoms in the $F=2$ hyperfine level at a temperature of $(16.8\pm0.7)\,\mu $K, and a phase space density $PSD_B\equiv n \lambda_{dB}^3=\left(4.5\pm 0.6\right)\times 10^{-7}$, where $n$ is the peak spatial density and $\lambda_{dB}=h/\sqrt{2 \pi m k_B T}$ the thermal de Broglie wavelength. These values represent our reference for a comparison with the results obtained with the grey molasses. 

The number of atoms in the MOT is monitored by their fluorescence emission: once this has reached a fixed value, we switch off the MOT magnetic fields and start the molasses.
To assess the efficiency, immediately after the molasses we measure: the number of remaining atoms, the temperature and the size of the sample. To measure the temperature, atoms are let free to expand for a certain time-of-flight (TOF), then we switch on the MOT laser
beams and acquire fluorescence images of the atomic cloud on a CCD camera for different TOF values (more details in Methods). 

\begin{figure}[h!]
\centering
\includegraphics[width=0.5\linewidth]{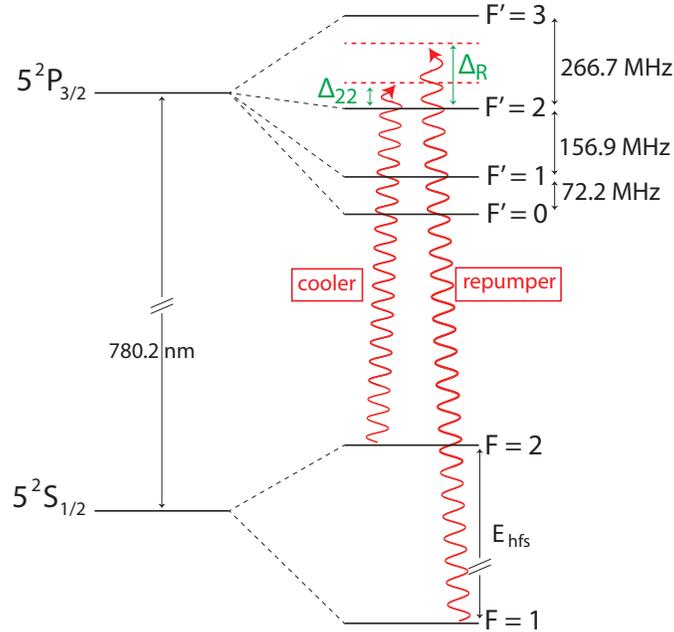}
\captionsetup{justification=justified}
\caption{ \textbf{Scheme of the energy levels of the $D_2$ transition in $^{87}$Rb.} Red arrows represent the two frequencies employed for the grey molasses; cooler(repumper) is blue-detuned with respect to the $F=2(1)\rightarrow F'=2$ transition, with detuning $\Delta_{22}$($\Delta_R$).
$\Ehfs$ denotes the ground-state hyperfine splitting, equal to $h\times 6834.683$ MHz. All other hyperfine splittings are indicated with the corresponding frequencies.}
\label{fig:LevelScheme}
\end{figure}

In most laser cooling experiments the cooler and the repumper lights are typically obtained from two distinct, not phase-coherent, laser sources. In the case of \rb, the cooler light is quasi-resonant on the $F=2\rightarrow F'=3$ transition and the repumper light is resonant on the $F=1\rightarrow F'=2$ (see
\fref{fig:LevelScheme} for the atomic levels structure of \rb). 
For grey molasses cooling enhanced by
$\Lambda$-configuration, the phase-coherence between cooler and repumper is necessary to preserve the linear superpositions of $F=1$ and $F=2$ sublevels which constitute the (quasi-)dark states (see in section "Adiabatic energy levels: dark states"). Indeed, we elucidate the relevance of phase-coherence between the two laser fields on the efficiency of grey molasses cooling by using alternatively a phase-coherent or a phase-incoherent repumper.
The phase-coherent repumper is obtained from a sideband of the cooler,
coherently generated by an Electro-Optical Modulator
(EOM) with a frequency shift $\Delta_{RC}/\left(2\pi\right)$ from the
carrier. In \fref{fig:LevelScheme} we display a schematic representation of \rb\ levels and the frequencies used in the molasses phase is
shown: cooler(repumper) is blue-detuned with respect to the
$F=2(1)\rightarrow F'=2$ transition, with detuning
$\Delta_{22}(\Delta_{R})$. Their frequency difference, as set by the radio-frequency driving the EOM, is
$\Delta_{RC}=\Delta_{R}-\Delta_{22}+\Ehfs/\hbar$ ($\Ehfs/h= 6834.68\,\text{MHz}$ is the energy splitting between the two hyperfine states $F=1$ and $F=2$ of the atomic ground state $5^{2}\text{S}_{1/2}$).

In order to fully characterise and optimise the grey molasses, we individually vary the repumper intensity, the molasses duration and the light detunings, and we measure the temperature ($T$), the sizes of the cloud ($\sigma_{x,z}$) and the number of atoms remaining after the grey molasses ($N$). From the measured data, we extract and examine also the $PSD$ of the sample, normalized to the value $PSD_{B}$ obtained with the bright molasses. 

\subsection*{Repumper intensity}

We start by describing the effect of varying the ratio between repumper and
cooler intensities $I_R/I_C$, which determines the superposition of states composing the dark-state. We fix the  molasses duration $\Delta t= 3\,$ms,  the optical intensity to $\sim 6 I_s$ for each beam and the detunings 
  $\Delta_R\simeq\Delta_{22}= 5\Gamma$
($I_s=1.67$ mW/cm$^2$ and $\Gamma = 2\pi \times 6.065$ MHz denote the saturation intensity and linewidth of the $D_2$ transition, respectively).
 We set the EOM sidebands frequency at
$\Delta_{RC}/(2\pi)=6834.6$ MHz, close to the hyperfine splitting $\Ehfs/h$,  
thus resembling a $\Lambda$-system, and
vary their amplitude, hence $I_R/I_C$, by adjusting the
radio-frequency power driving the EOM. For each value we measure the temperature $T$, the fraction of remaining atoms $N/N_{MOT}$ and the normalized phase space density $PSD/PSD_{B}$ extracted from the data.

\begin{figure}[h!]
\centering
\includegraphics[width=0.51\linewidth]{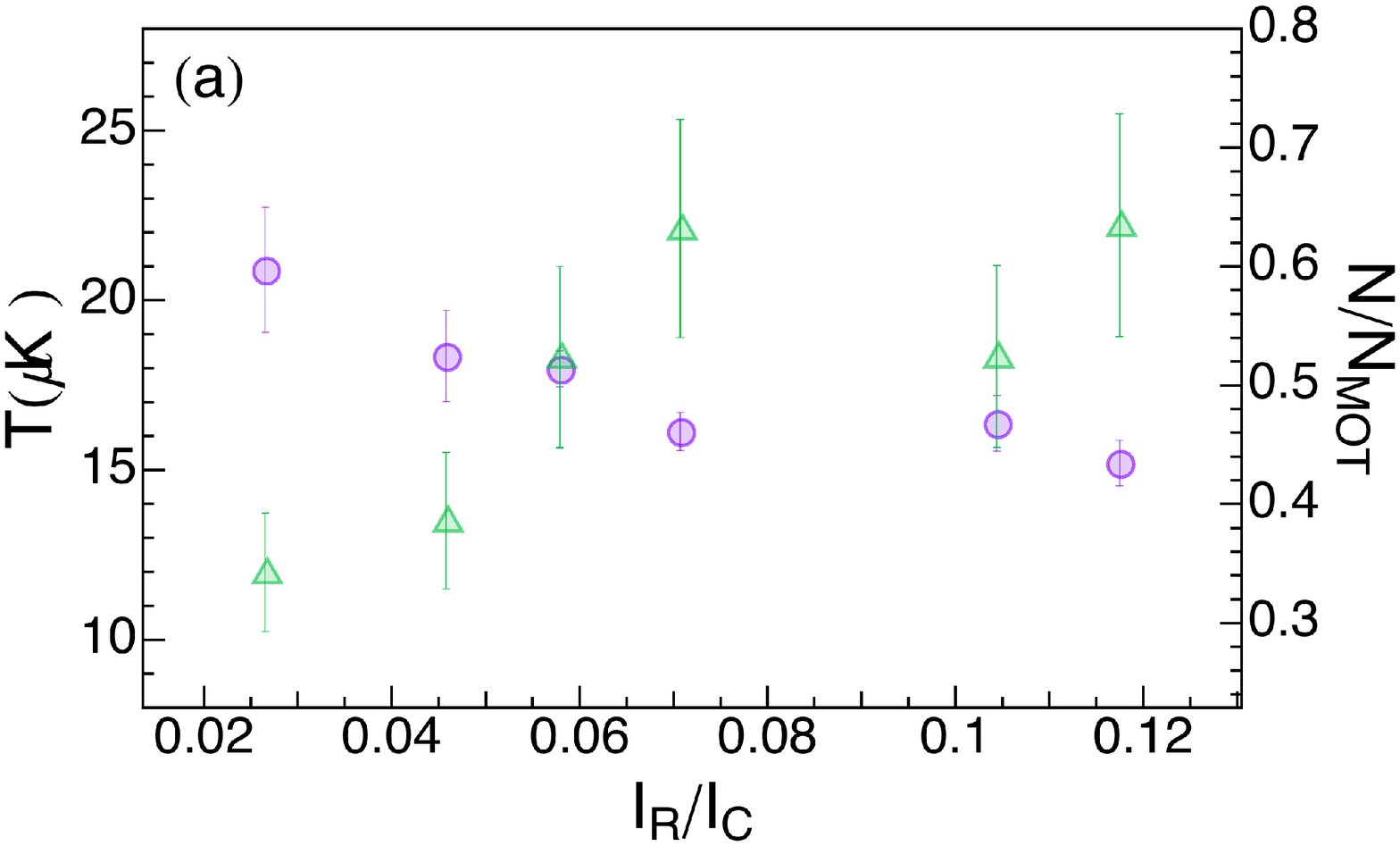}\hfill
\includegraphics[width=0.47\linewidth]{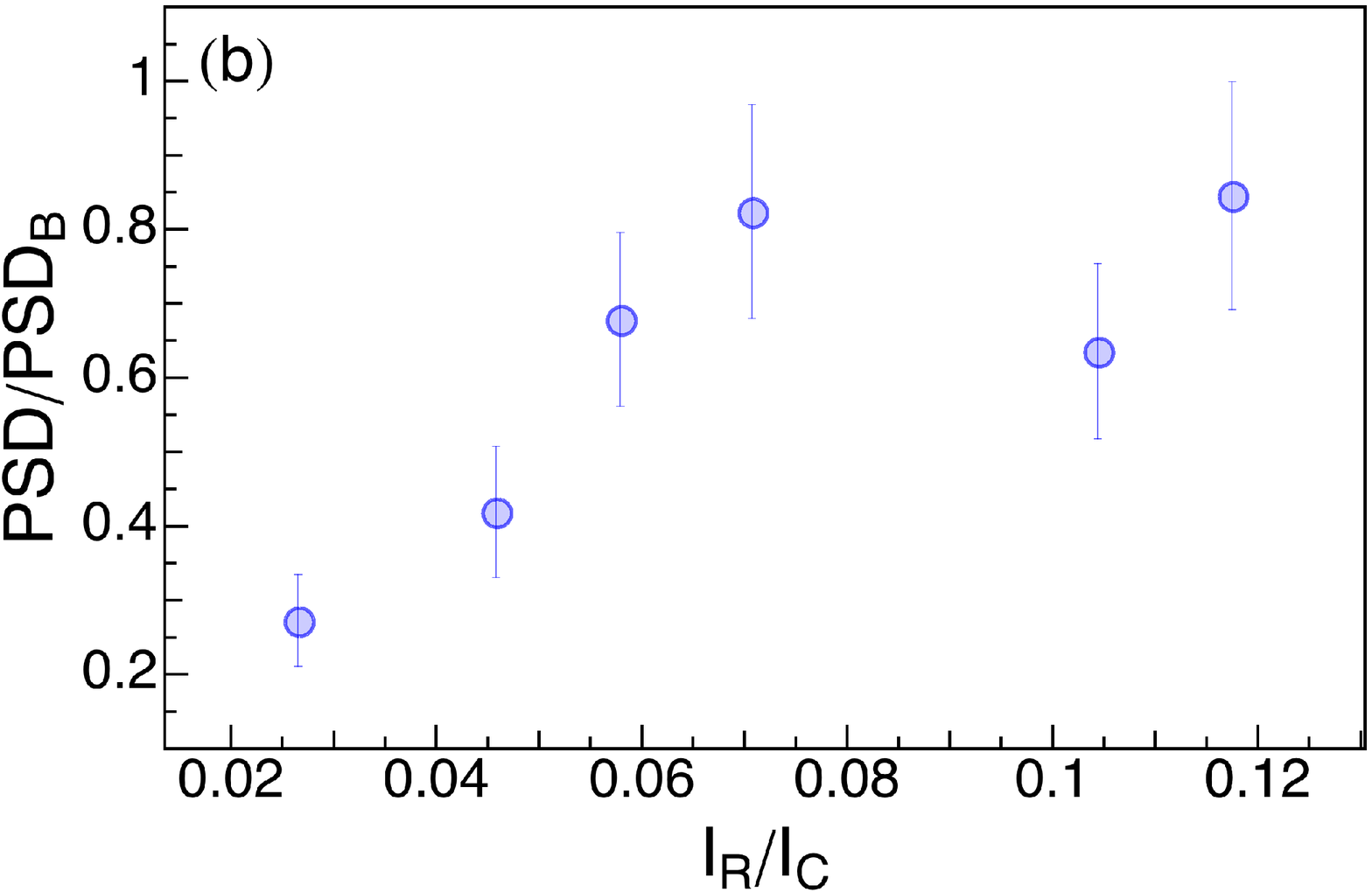}
\captionsetup{justification=justified}
\caption{\textbf{Effect of repumper intensity. } (a): Temperature $T$ (purple circles) and ratio between atom number
  after and before the molasses $N/N_{MOT}$ (green triangles) as a function of the intensity ratio $I_R/I_C$ between repumper and
  cooler light; (b): $PSD$ measured after grey molasses normalized to the value of the bright molasses ($PSD_B$). 
}
\label{fig:TandNN0_vsRepFraction}
\end{figure}

Data in \fref{fig:TandNN0_vsRepFraction}(a) show that 
$I_R/I_C\simeq0.07$ is sufficient to have both minimum temperature and maximum atom number while the 
$PSD$ in \fref{fig:TandNN0_vsRepFraction}(b) saturates for higher values of the repumper intensity fraction. A relatively small value of the repumper intensity could be convenient as it simplifies the requirements on the power of the sidebands produced by the EOM. 

We point out that the lowest temperature in the \fref{fig:TandNN0_vsRepFraction}(a) is larger than the minimum value reported below because
these data were taken with imperfect cancellation of the residual magnetic field, a critical step to optimize the performance of the grey molasses. 
Data presented hereafter were instead obtained, after proper magnetic field compensation, with $I_R/I_C$ set to 0.08. 

\subsection*{Time duration}

The time duration of the grey molasses
is crucial for its 
efficiency. Ideally, for perfect molasses, a long duration reduces the final temperature. However, due to the absence of any spatial trapping, the density of the sample decreases as the atoms diffuse and expand. As a consequence, if the parameter to be maximized is the $PSD$, a compromise arises 
between lower temperature and higher density. 

\begin{figure}[h!]
\centering
\includegraphics[width=0.51\linewidth]{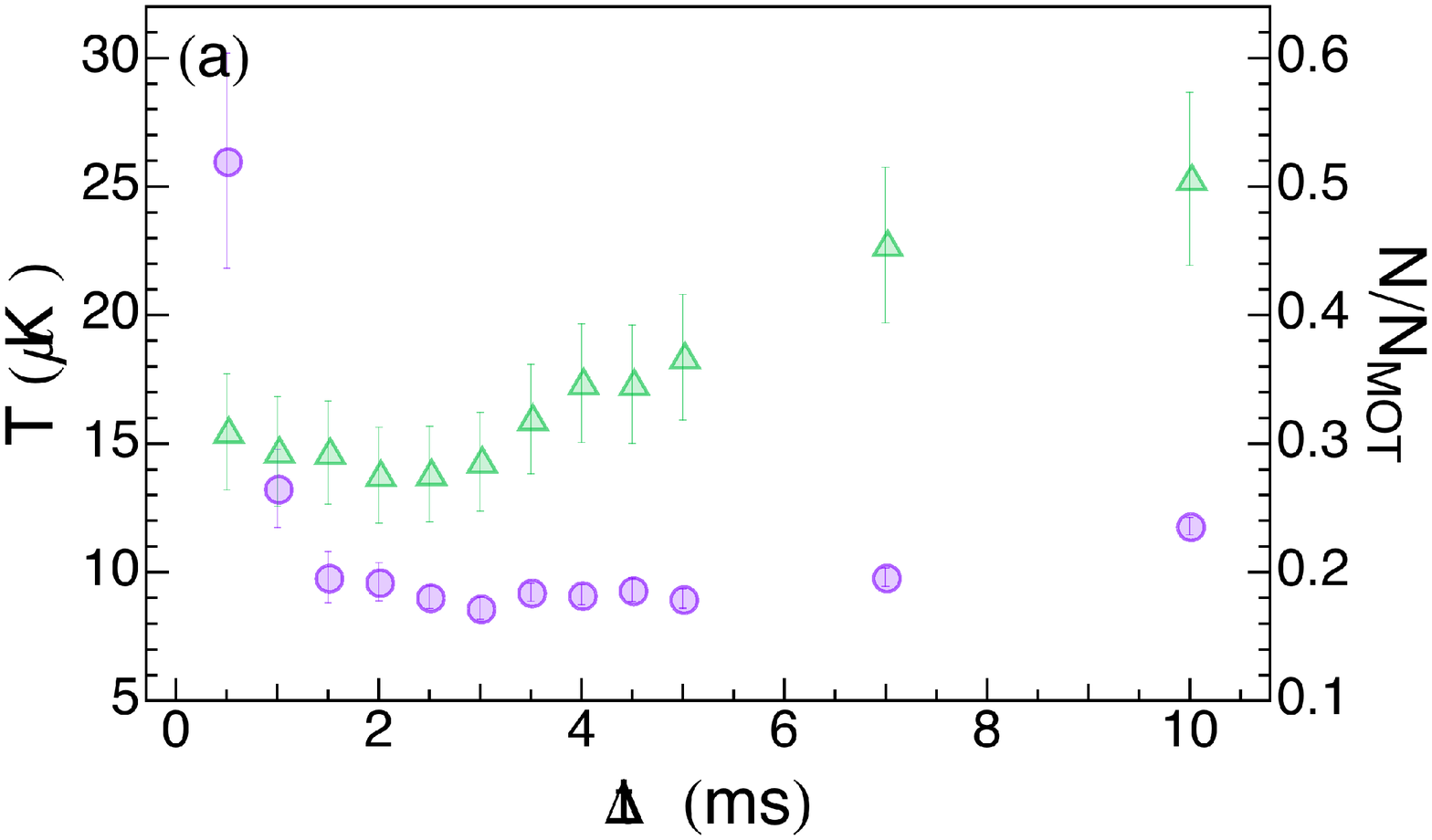}\hfill
\includegraphics[width=0.47\linewidth]{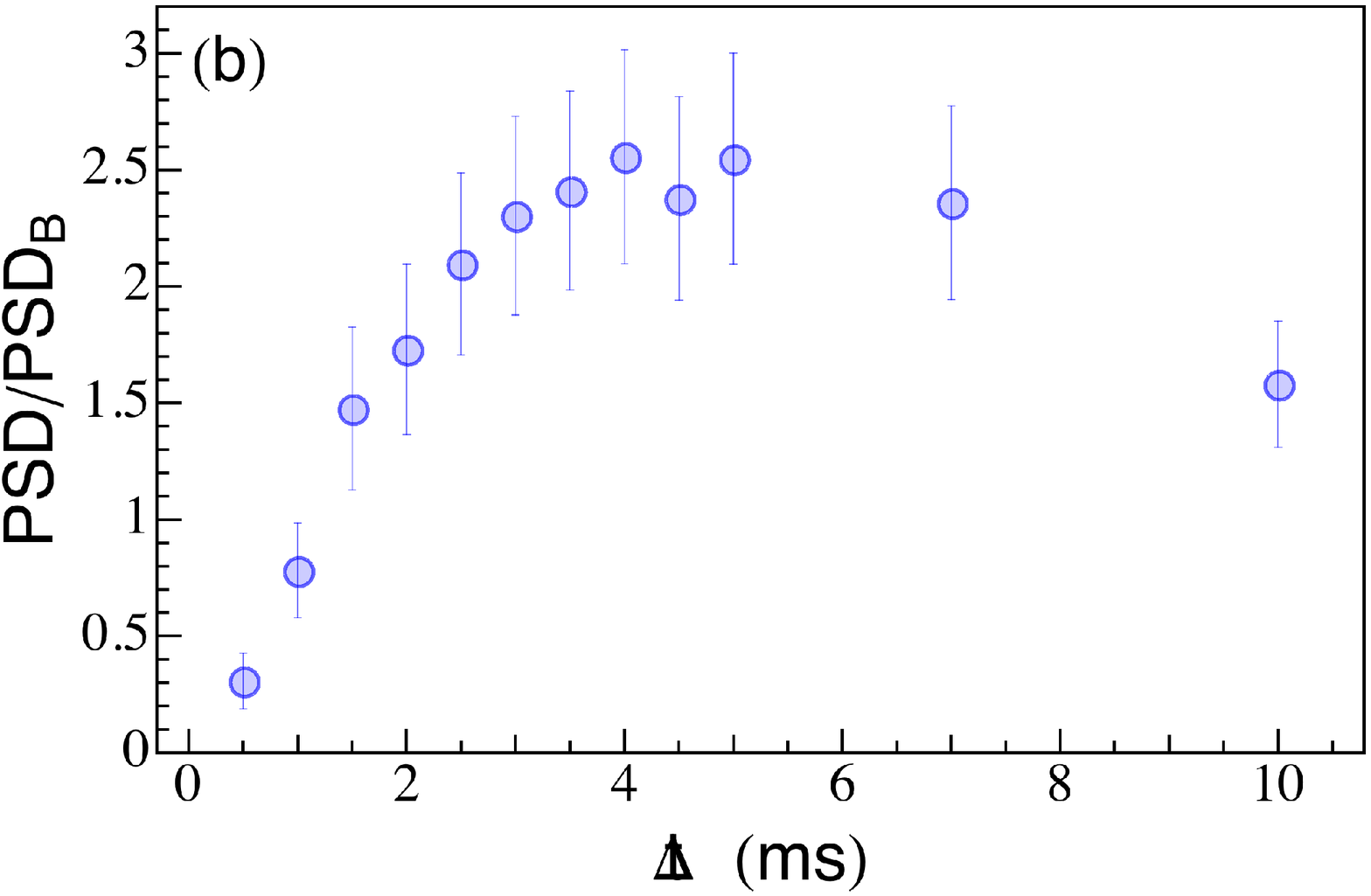}
\captionsetup{justification=justified}
\caption{\textbf{Optimization of molasses duration. } (a): Temperature $T$ (purple circles), fraction of remaining atoms $N/N_{MOT}$ (green triangles) and (b):  normalized phase space density $PSD/PSD_{B}$ as a function of the duration $\Delta t$ of the grey molasses.}
\label{fig:TandNN0_vsmoltime}
\end{figure}

In order to find its optimal value, we vary the molasses duration $\Delta t$ at constant detunings $\Delta_R\simeq\Delta_{22}=5 \Gamma$ and $\Delta_{RC}/(2\pi)=6834.6\,\text{MHz}$. 
\fref{fig:TandNN0_vsmoltime}(a) shows that the temperature initially decreases
with $\Delta t$, reaching its minimum value
$T_{min}=\left(8.6\pm0.4\right)\mu\text{K}$ after 3 ms of molasses, then it
flattens and slightly increases for longer times (at 10 ms we measure
$12\,\mu$K). Conversely, we find that longer molasses capture more atoms. 
To choose the optimal duration, we plot the $PSD$ in 
\fref{fig:TandNN0_vsmoltime}(b). Within the error bars, 
the $PSD$ is
maximum for time durations from $\Delta t=3\,\text{ms}$ to
$\Delta t=7\,\text{ms}$, and 
the temperature is minimum within the
whole range. Therefore, in the following, we fix the duration $\Delta t$ to 3 ms.

\subsection*{Phase-coherence and detunings} 

As expected, the detuning of the cooler with respect to the open transition $F=2\rightarrow F'=2$ (and as a consequence the detuning of the repumper with respect to the $F=1\rightarrow F'=2$ transition) considerably influences the efficiency of the grey molasses. In addition, coherence between the two frequencies is fundamental to preserve the superpositions of atomic sublevels composing the dark states. For this reason, in this section we investigate in detail the dependence of the efficiency of grey molasses on the absolute detunings $\Delta_{22}$ and $\Delta_{R}$ both in the case of coherent and incoherent light, keeping their relative detuning $\Delta_{RC}$ fixed at $\left(2\pi\right)\times 6834.6\,\text{MHz}$. The detunings $\Delta_{22}$ and $\Delta_{R}$ therefore take almost the same value; for simplicity the data in the plots are reported as a function of $\Delta_{22}$. 

\begin{figure}[h!]
\centering
\includegraphics[width=0.49\linewidth]{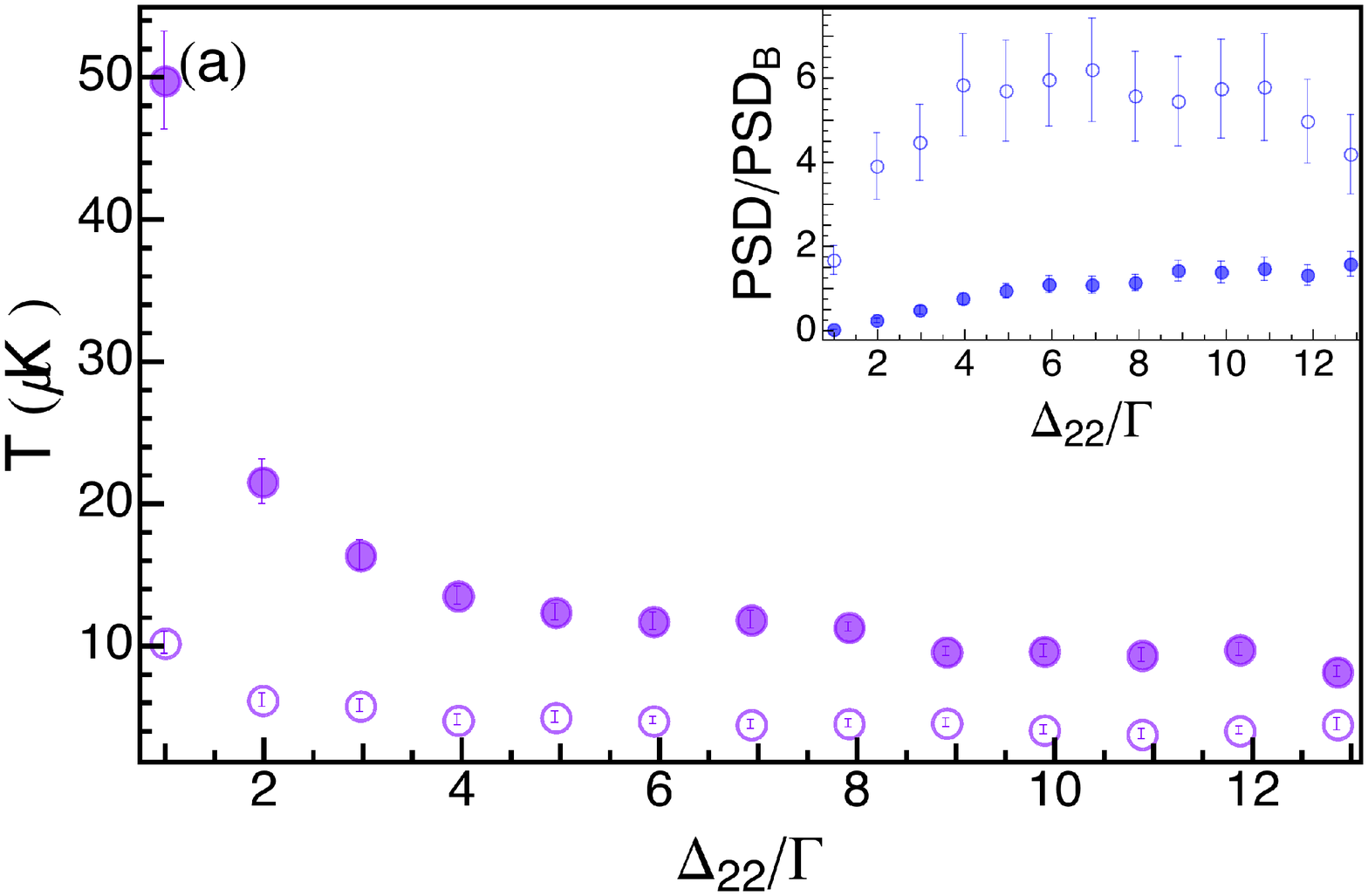}
\hfill
\includegraphics[width=0.49\linewidth]{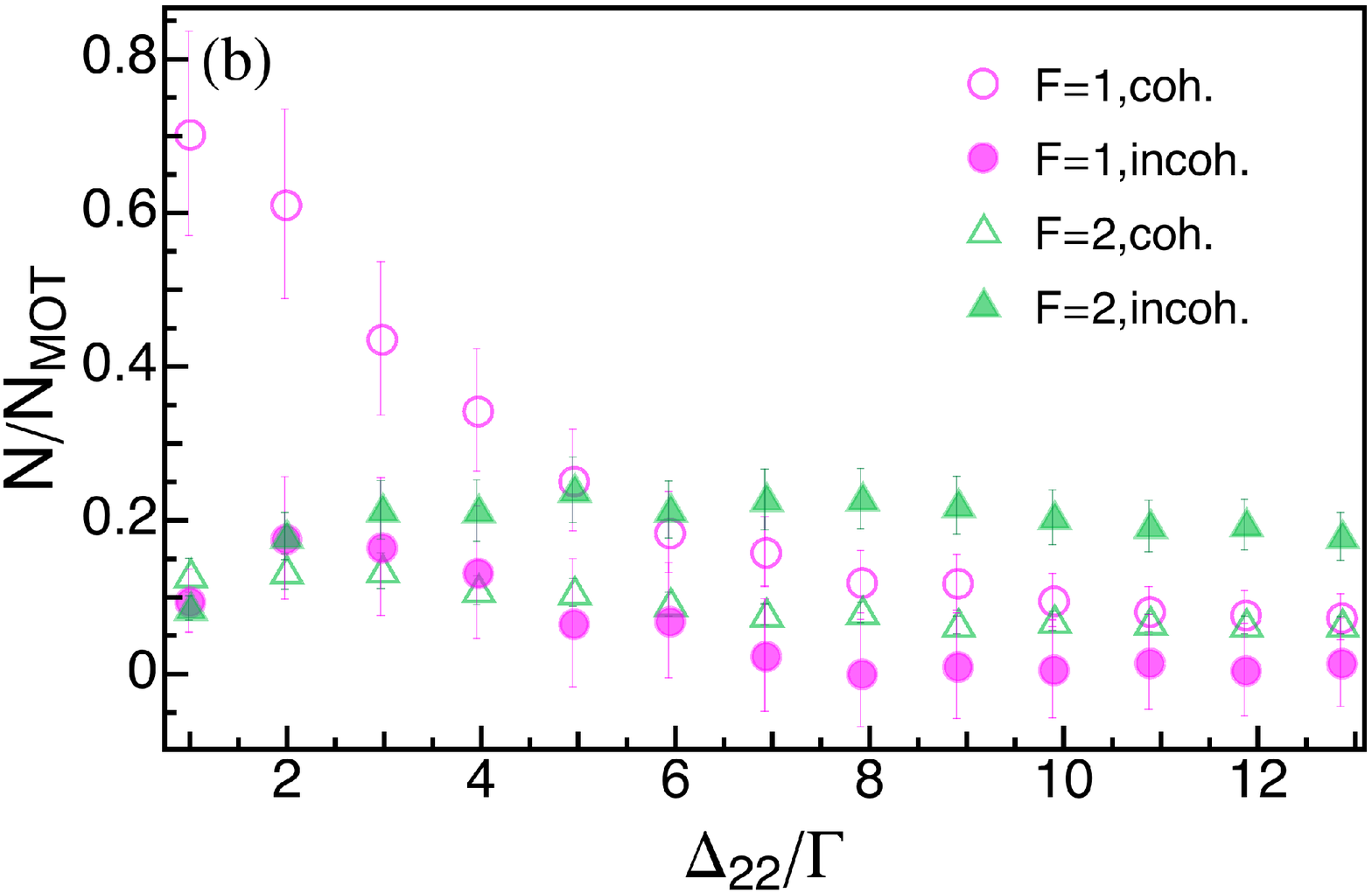}
\captionsetup{justification=justified}
\caption{\textbf{Influence of detuning: incoherent and coherent repumper. } (a) Temperature $T$ ($PSD/PSD_{B}$ is shown in the inset) as a function of the cooler and rempumper detuning. We report both the case of coherent (empty points) and incoherent repumper (filled points). (b) Fractional number of remaining atoms $N/N_{MOT}$ in $F=1$ (pink circles) and $F=2$ (green triangles); empty (filled) points refer to the case of coherent (incoherent) repumper.}
\label{fig:TNN0PSD_vsOVdet}
\end{figure}

In \fref{fig:TNN0PSD_vsOVdet}(a) we compare two datasets: one refers
to the grey molasses performed using the coherent repumper light (empty points in the graph),
while the other is obtained using the repumper light delivered by
another distinct laser source (filled points in the graph), i.e. incoherent with respect to the
cooler. In both cases, the temperature decreases as we increase the
detuning, but the use of the coherent repumper light clearly provides
lower temperatures in the whole range of detunings. The trend of $PSD/PSD_B$ is reported in the inset. 

In \fref{fig:TNN0PSD_vsOVdet}(b) the same comparison is reported for
$N/N_{MOT}$, where, we remind, $N_{MOT}$ is the total number of atoms measured at the end of the MOT phase. Here, both for the case of coherent and incoherent light, we compare the number of remaining atoms in the $F=1$ and $F=2$ ground hyperfine levels. As a matter of fact, also the knowledge of the final state of the cooled atoms is important, in particular once the atoms are subsequently transferred, and further cooled, in state-dependent traps. Besides, the final state of the atoms crucially depends on the mechanism underlying the grey molasses, and provides an effective probe on the cooling operation. 

In the case of coherent repumper light, for small detunings almost 80\% of the atoms are cooled in the grey molasses and they mostly occupy the $F=1$ hyperfine state. Increasing the detuning, the total number of atoms drastically decreases and their population is equally distributed among the two hyperfine states $F=1$ and $F=2$. With incoherent repumper light, instead the grey molasses always captures a small fraction of the MOT atoms; also, the relative population in the two hyperfine levels is almost equal for small detunings while for higher ones the atoms are mostly pumped in the $F=2$ state.
Close to the resonance, the observed trend is consistent with the expected behavior for our experimental parameters, i.e. the fraction of atoms effectively pumped in $F=1$ almost reflects the cooler/repumper intensity ratio. The comparison provided here is a stark evidence that coherent evolution within the ground hyperfine manifold enhances the cooling efficiency of the grey molasses and the accumulation of atoms in $F=1$ level.

The best compromise between low temperatures and large atom numbers is
found looking at the $PSD$ in the inset of \fref{fig:TNN0PSD_vsOVdet}(a): we report the case of coherent light (empty points) and incoherent (filled points). Actually, in the case of coherent light,
the $PSD$ levels to its maximum value in the broad range of detunings from $\Delta_{22} \simeq 4\Gamma$ to
$\simeq 11\Gamma$, and it is much higher than in the incoherent case within the whole range of detunings. 

\subsection*{Raman detuning}

\begin{figure}[h!]
\centering
\includegraphics[width=0.49\linewidth]{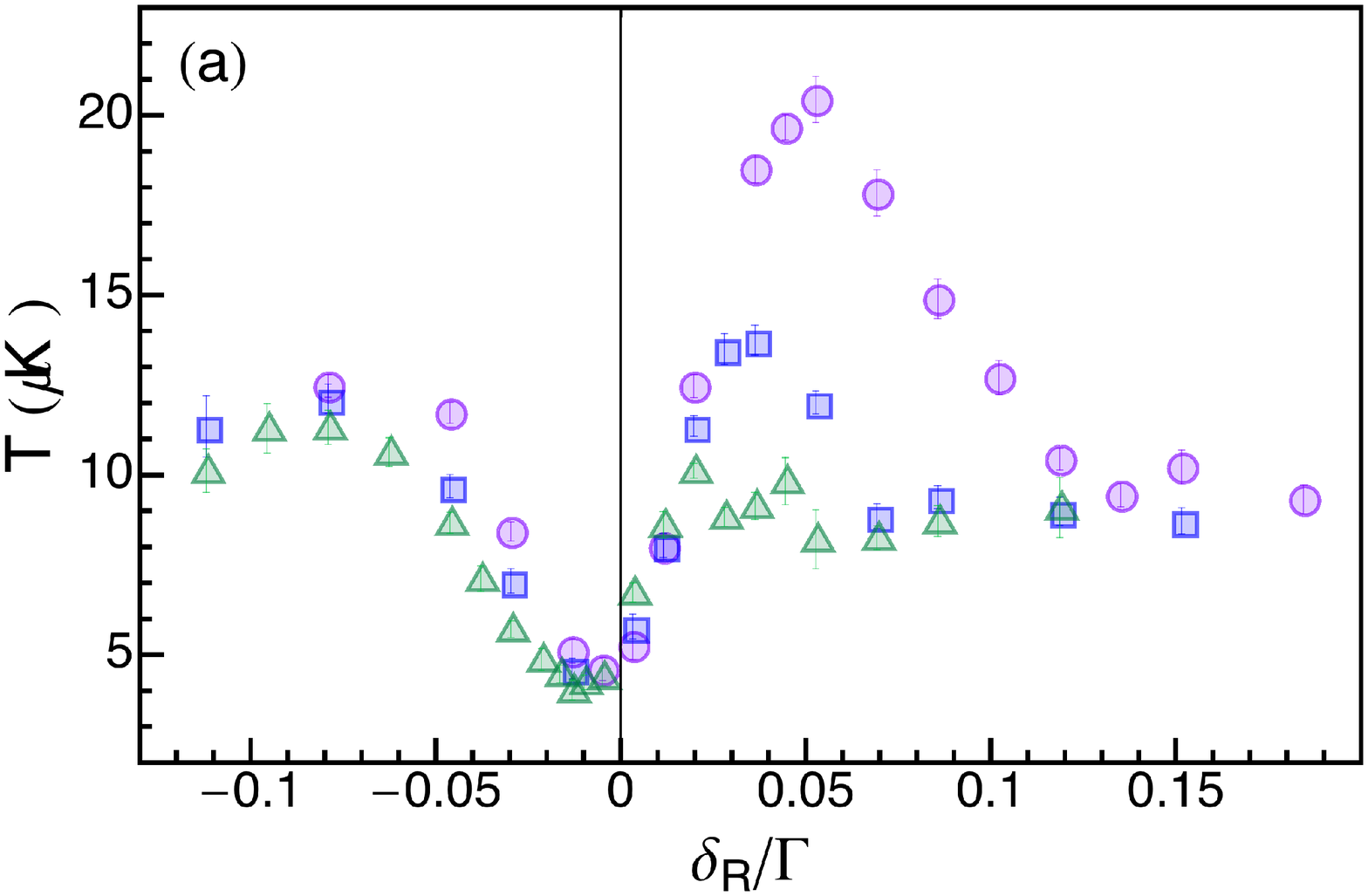}
\hfill
\includegraphics[width=0.49\linewidth]{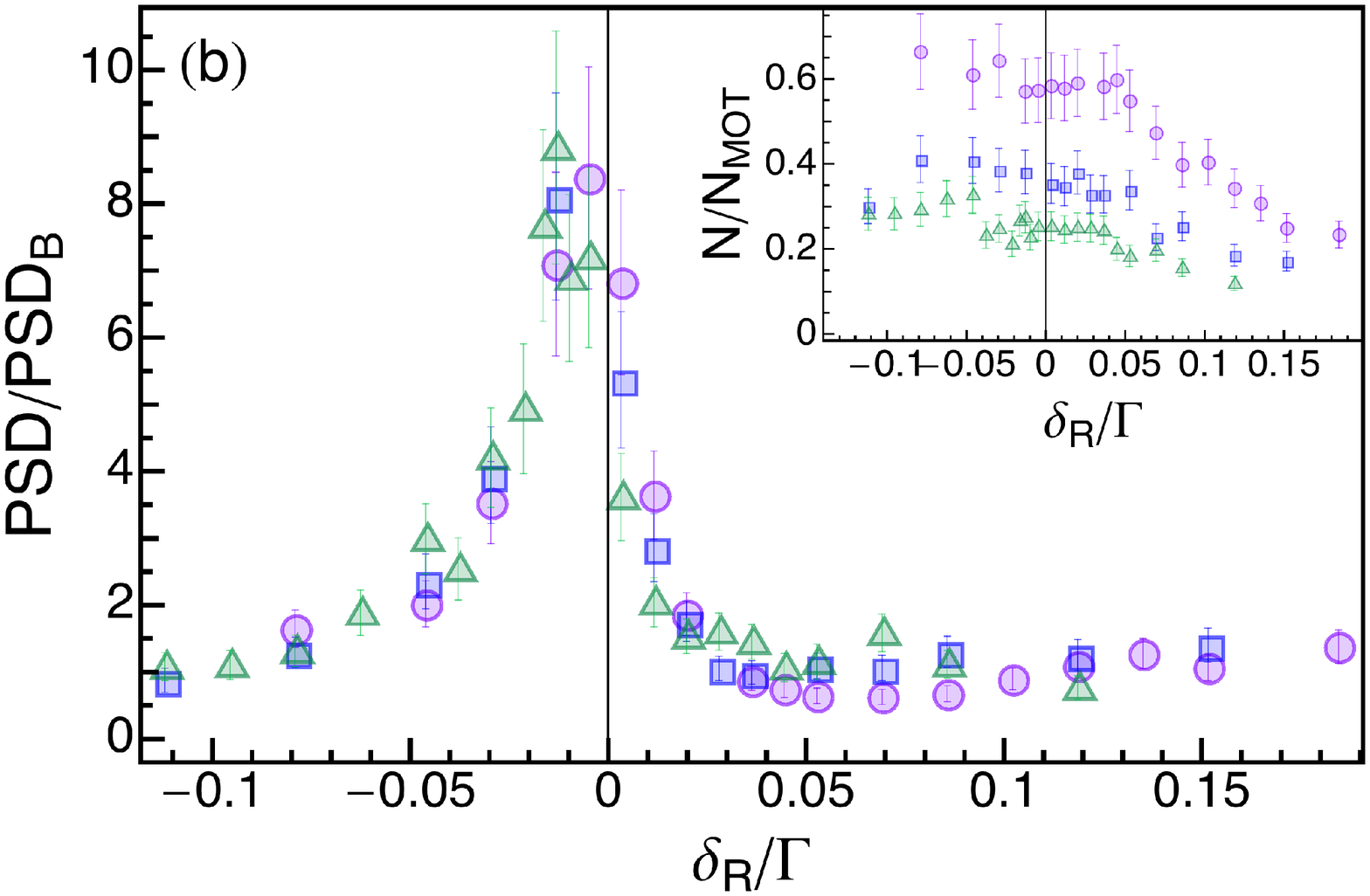}
\captionsetup{justification=justified}
\caption{\textbf{Raman detuning. }(a): Temperature $T$ and (b): normalized phase space density $PSD/PSD_{B}$ ($N/N_{MOT}$ is shown in the inset) as a function of the Raman detuning $\delta_{R}$. In both graphs, we show three different
  datasets: $\Delta_{22}=5\,\Gamma$ (pink circles), $\Delta_{22}=8\,\Gamma$
  (blue squares) and $\Delta_{22}=12\,\Gamma$ (green triangles).}
\label{fig:TNN0_vsRAMANdet}
\end{figure}

In order to optimize the relative detuning between the coherent repumper and the cooler we vary the EOM frequency, for three different values of $\Delta_{22}$: $5\,\Gamma$, $8\,\Gamma$ and
$12\,\Gamma$. In this section, we 
define the Raman detuning $\delta_{R}\equiv\Delta_{RC}-\Ehfs /\hbar$ as the detuning of the repumper light with respect to the Raman condition of the $\Lambda$-configuration. The experimental data in \fref{fig:TNN0_vsRAMANdet} are reported as a function of this quantity. 

The temperature measured after the grey molasses is reported in
\fref{fig:TNN0_vsRAMANdet}(a): for each value of
$\Delta_{22}$, the minimum temperature is obtained for
$\delta_R\simeq -0.01 \Gamma$, slightly below the Raman condition $\delta_R=0$,
similar to what has been observed in earlier experiments \cite{colzi_grey_2016}. Despite a careful compensation of stray magnetic fields, we could not eliminate the residual shift with respect to the Raman condition. Consistently with the data shown in the previous section, we see that the minimum temperature is approximately the same for the three values of $\Delta_{22}$. In the inset of \fref{fig:TNN0_vsRAMANdet}(b) we report the number of remaining atoms; we observe that in the whole range of $\delta_{R}$ the higher is $\Delta_{22}$, the smaller is $N/N_{MOT}$, in agrement with the trend observed in \fref{fig:TNN0PSD_vsOVdet}(b). 

The corresponding values obtained for $PSD/PSD_B$ are reported in \fref{fig:TNN0_vsRAMANdet}(b); we observe a clear peak for $\delta_R\simeq -0.01 \Gamma$, and the maximum value of the $PSD$ is the same within the error bar for the three different
$\Delta_{22}$ values, as shown in the summary table \ref{tableRaman}.
\begin{table}
\centering
\begin{tabular}{|c|c|c|c|}
\hline
$\Delta_{22}/\Gamma$ & T $\left[\mu\text{K}\right]$ & $N/N_{MOT}$ & $PSD/PSD_{B}$ \\
\hline
5 & $\left(5.1\pm0.3\right)$ & $\left(0.57\pm0.08\right)$ & $\left(7.1\pm0.9\right)$ \\
8 & $\left(4.6\pm0.3\right)$ & $\left(0.38\pm0.05\right)$ & $\left(8.0\pm0.9\right)$ \\
12 & $\left(4.0\pm0.3\right)$ & $\left(0.28\pm0.04\right)$ & $\left(8.9\pm0.9\right)$ \\
\hline
\end{tabular}
\caption{Minimum temperature $T$, fraction of remaining atoms $N/N_{MOT}$ and normalized phase space density $PSD/PSD_B$ measured for $\delta_R=-0.01\,\Gamma$.}
\label{tableRaman}
\end{table}

However, a notable difference arises for positive values of $\delta_R$ when we compare the datasets for $\Delta_{22}/\Gamma=5,\,8,\,12$. Here, as
expected, cooling is less efficient \cite{grier_lambda-enhanced_2013} and the
final temperature features a Fano profile, which however gets lower for larger
detunings, to the point that for $\Delta_{22}=12\Gamma$ it is no longer
visible.
The reason of this behaviour might lie in the structure of $^{87}$Rb atomic
levels: the energy separation between $F'=2$ and $F'=3$ levels is $44\,\Gamma$, thus for high values of $\Delta_{22}$ the upper level might start to play a
role. A deeper understanding of this effect could further elucidate how grey molasses works in the presence of a richer level structure, but requires additional investigations.

\section*{Adiabatic energy levels: dark states}

It was earlier recognized that physical insight into this cooling mechanism is gained through a 1-dimensional model of
$\Lambda$-enhanced grey molasses \cite{weidemuller_novel_1994,grier_lambda-enhanced_2013}, that takes into account the variations in space of the levels of the full Hamiltonian dressed by the laser fields.  
Thus we calculate here the position-dependent energy levels by numerical diagonalization of the Hamiltonian
\begin{equation}
H=\sum_{jFm} E_{jFm} |j, F, m\rangle \langle j, F, m| - \frac{\hbar}{2}[(\Omega_Re^{-i\omega_R t}+\Omega_C e^{-i\omega_C t})(\hat{y}e^{ikz}+\hat{x}e^{-ikz})\vec{a} + h.c.]   
\end{equation}
taking into account the full hyperfine structure of both the lower
$5 ^2 S_{1/2}$ and upper $5 ^2 P_{3/2}$ electronic levels of the $D_2$
transition: $E_{jFm}$ denote the energies of the ground ($j=1/2$) and excited ($j=3/2$) hyperfine manifolds, with the definition $E_{1/2, 1, m}=0$; $\Omega_{R(C)} \equiv \Gamma \sqrt{I_{R(C)}/2 I_S}$ is the repumper (cooler) Rabi frequency in terms of the saturation intensity $I_S=1.67$ mW/cm$^2$ and the excited state linewidth $\Gamma/(2\pi)=$ 6.065 MHz, $\vec{a}$ are the raising operators of atomic levels whose
matrix elements are the 6-j Wigner coefficient and, finally, $\omega_{R(C)}$ the repumper (cooler) angular frequency. We consider a configuration with
two counter-propagating beams of orthogonal linear polarizations
(lin$\perp$lin). Each beam carries the repumper and cooler frequency
$\omega_R , \omega_C$, with Rabi frequencies corresponding to the total intensities
used in the experiment in all six beams, namely
$\Omega_C=4.2\,\Gamma, \Omega_R=1.2\,\Gamma$.
First, we neglect the coupling of the cooler (repumper) with the
$F=1(2)\rightarrow F'$ transitions, due to very large detuning
($\sim 10^3 \Gamma$).
Then, we apply the unitary transformation 
$$U = P_1 + \exp[i (\omega_R-\omega_C) t]P_2 +\exp[i \omega_R t]P_e$$
where $P_1, P_2, P_e$ are the projectors on the ground lower
$\{|j=1/2,F=1,m\rangle\}$, ground upper $\{|j=1/2,F=2,m\rangle\}$, and
electronic excited $\{|j=3/2,F',m'\rangle\}$ hyperfine levels,
respectively. Under the above unitary transformation, the Hamiltonian is
modified $H'\equiv UHU^\dagger + i (\partial_t U) U^\dagger$: (i) the
time-dependence of the Rabi terms is canceled, (ii) the energy levels are
shifted
$E'_{1/2,1,m}=0,\quad E'_{1/2,2,m} = \Delta_R - \Delta_C,\quad E'_{3/2,F,m} =
E_{3/2,F,m}- \omega_R$.

\begin{figure}[h!]
\centering
\includegraphics[width=0.48\linewidth]{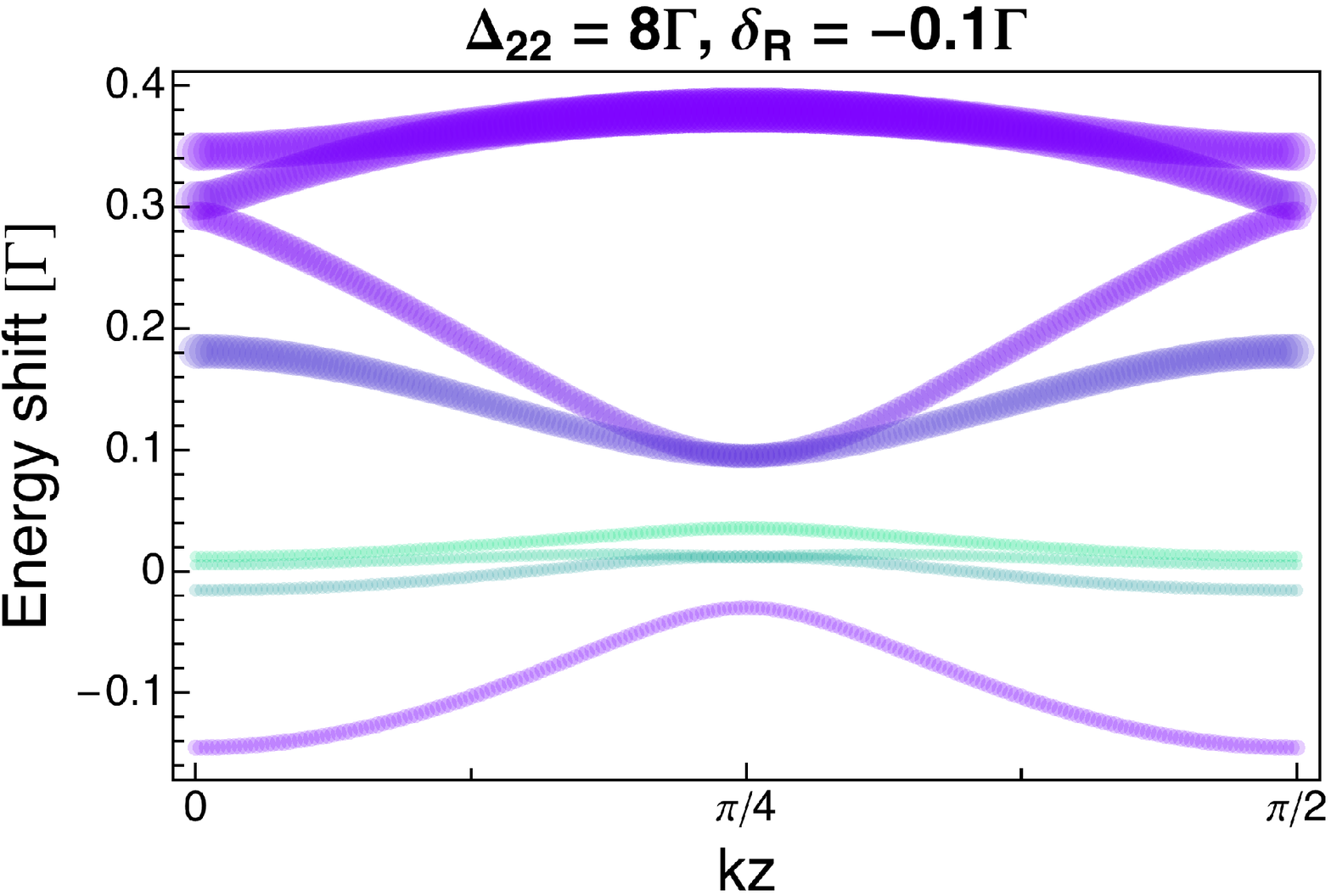}\hfill
\includegraphics[width=0.48\linewidth]{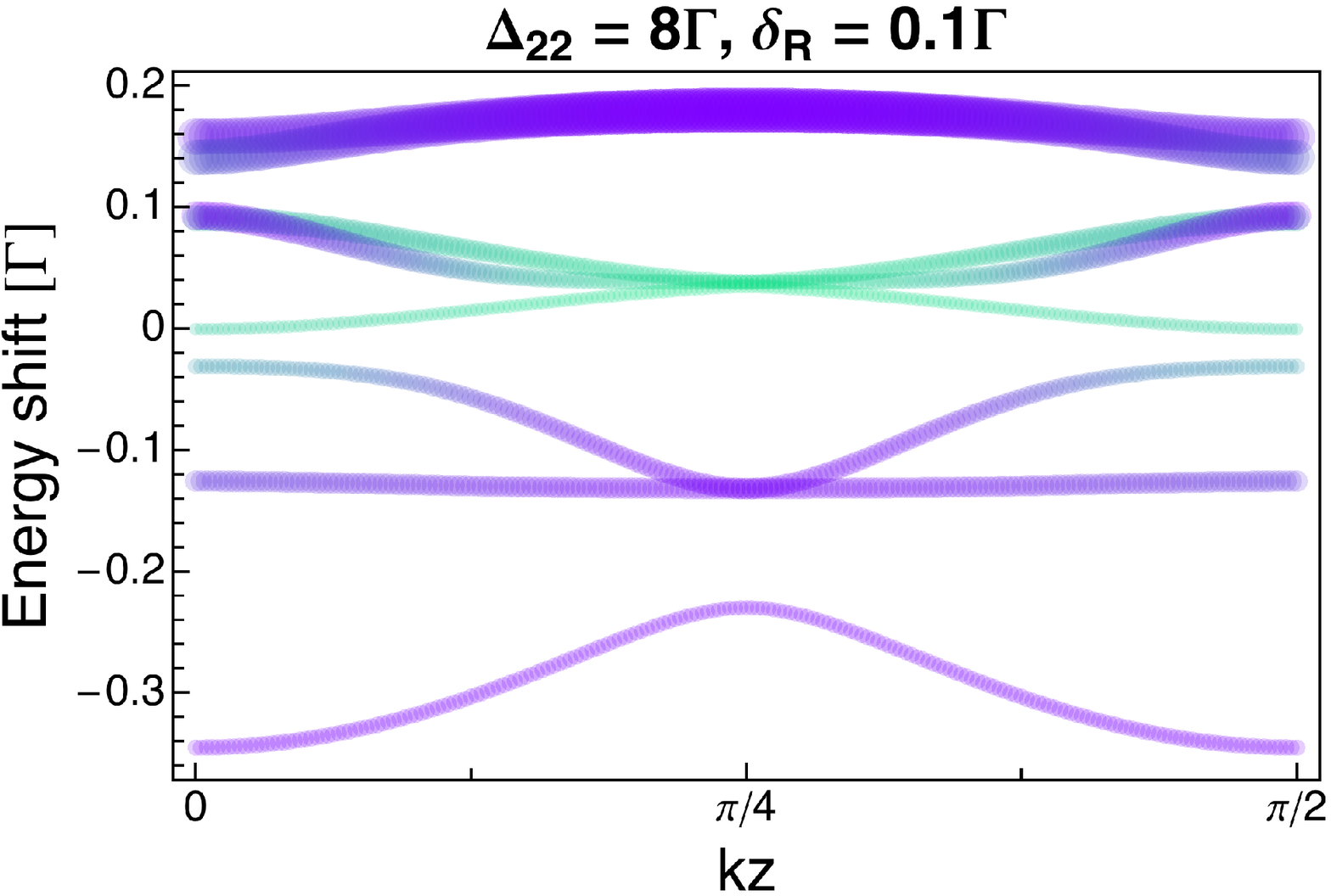}
\captionsetup{justification=justified}
\caption{\textbf{Energy of low-scattering states as a function of position.} Purple(green) lines are states with dominant weight in $F=1(2)$ level; for each state considered, the values plotted in graph are the energy shifts with respect to the corresponding level energy in the absence of the light. The line thickness is proportional to position-dependent scattering rate $\gamma'$ defined in text. (a): Energies calculated for $\delta_{R}=-0.1\Gamma$; (b): energies calculated for $\delta_{R}=+0.1\Gamma$.}
\label{fig:darkstates}
\end{figure}

Figure \ref{fig:darkstates} shows the position-dependent eigenvalues of $H'$. 
For each state $|\psi_j\rangle$, the line-thickness encodes the scattering rate $\gamma_j'=\Gamma \langle \psi_j | P_e|\psi_j\rangle$, as these are the interesting states for the grey molasses mechanism. 
We plot only states with $\gamma_j'/\Gamma<0.5$, whose population is
predominantly in the two ground hyperfine manifolds. It is clear that the low
scattering states are mainly in the $F=1$ ground level and that $F=2$ states are
generally broader, as expected from the relative magnitude of the Rabi
frequencies, $\Omega_C>\Omega_R$. We also notice that, for negative Raman
detuning $\delta_R=-0.1\Gamma$ (see \fref{fig:darkstates}(a)) the level
configuration favors cooling as several bright states lie at higher energy than
the low-scattering (narrow), predominantly $F=1$,  states. Conversely, for positive Raman
detuning $\delta_R=0.1\Gamma$ (see \fref{fig:darkstates}(b)), the predominantly $F=1$ levels are visibly more
mixed with the other levels.

We point out that the grey molasses cooling occurs because, thanks to their motion, atoms in dark states still have a
finite probability amplitude of undergoing diabatic transition to a different adiabatic dressed state. Quite reasonably, the probability amplitude of these Landau-Zener (LZ) processes is proportional to the atomic velocity and is larger at the locations where the dressed states get closer in energy (avoided crossings).
If, following the diabatic transition, an atom ends in a bright state it faces
two possibilities: either climbing or sloping down the light-shift potential as
it moves away from the LZ location, the former (latter) leading to loss (gain)
of kinetic energy, i.e. cooling (heating).  Obviously the bright state energies
are periodic in space, thus the cumulative variation of kinetic energy vanishes
when the atom travels over one period distance. But the lifetime of bright
states is limited by optical pumping, so that the average variation of kinetic
energy is determined by the dynamics immediately after the LZ transition.

More quantitatively, we calculate the diabatic
couplings $\sim v \langle\psi_j| \partial_z \psi_k\rangle$ for any given state
$|\psi_j\rangle$ to all other states $|\psi_k\rangle$, where $\partial_z$ and
$v$ denote the gradient and the velocity in our 1D model. The diabatic couplings
confirm the cooling scenario described above, showing for example that the
lowest-energy, predominantly $F=2$, level in \fref{fig:darkstates}(a) is not
coupled to the low-scattering, predominantly $F=1$, states.

We notice that clear differences in the landscape of the calculated
energy levels are not visible in the range of experimentally explored Raman detunings $\delta_R$ ($-0.05<\delta_{R}/\Gamma<0.05$) of \fref{fig:TNN0_vsRAMANdet}. Indeed in the calculations the values of Raman detuning results irrelevant if much smaller than the light shifts. Thus, in order to observe a difference in the level structure in the calculations we actually need to vary $\delta_{R}$ much more than in the experiment (see \fref{fig:darkstates}). The discrepancy could signal that we are performing the calculations with Rabi frequencies larger than the effective experimental values.
As a matter of fact the one-dimensional calculation performed considering lin$\perp$lin configuration cannot exactly account for the experimental configuration, which actually consists in three couple of counter-propagating beams, each couple made of circularly polarized $\sigma_{+}/\sigma_{-}$ beams. Nonetheless, as already demonstrated \cite{grier_lambda-enhanced_2013} the simplified scheme allows to capture the essential mechanism undergoing.

\section*{Conclusions}

In summary we have shown that efficient dark-state cooling can be achieved even on the $D_2$ transition for Rb atoms, thanks to the relatively large hyperfine separations of the upper level which make the $F=2\rightarrow F'=2$ open transition
sufficiently isolated from the closed $F=2\rightarrow F'=3$. 
We reported a thorough experimental characterisation of the grey molasses operation as a function of different experimental parameters such as the intensity of the repumper light, the time duration, and the frequencies of both cooler and repumper. Furthermore, we have pointed out the fundamental role of the phase coherence between the cooler and the repumper laser fields.

We find some interesting differences with respect to grey molasses on the $D_1$ transition, for example in the typical Fano profile shown in \fref{fig:TNN0_vsRAMANdet} where the high-temperature peaks arising for positive Raman detuning reduces at large detunings. Although the calculated energy levels, scattering rates and diabatic couplings provide useful hints, we lack a full explanation for this effect.

Our findings have practical consequences for experiments in cold atoms as they show that the $PSD$ can be increased with $\Lambda$-enhanced grey molasses without the drawback of an additional laser source on the $D_1$ transition. In addition, grey molasses achieve efficient optical pumping in $F=1$ level, which can be convenient for further experiments. In particular, for mixtures experiments where one species requires grey-molasses cooling, also on the other one it is useful to have it. 

\section*{Methods} \label{Sec:methods}

\subsection*{Experimental procedure: MOT and molasses} 

We start loading a 3D-MOT from an atomic beam similarly to that described in \cite{catani_intense_2006}. For the MOT loading, the repumper and cooler light are provided by two distinct diode lasers and overlapped on a single-mode polarisation-maintaining fiber, and they have detunings  $\Delta_{23}\simeq - 2.5\,\Gamma$ and $\Delta_R\simeq-0.5\,\Gamma$ with respect to the $F=2\rightarrow F'=3$ and $F=1\rightarrow F'=2$ transitions, respectively. The repumper intensity is about $5\%$ of the total 3D-MOT light. 
A pair of coaxial coils in anti-Helmholtz configuration generates the magnetic field
gradient, approximately 14 G/cm in the vertical direction and 7 G/cm along any
direction in the horizontal plane. The 3D-MOT consists of six independent laser
beams - two counter-propagating beams for each spatial direction - with an
intensity of about $6\,I_{\text{sat}}$ for each beam. We load the 3D-MOT for 5
to 7 seconds, up to a fixed number of $3\times10^8$ atoms. In
order to have the same number of atoms collected in the MOT for each
experimental run, we collect part of the fluorescence light emitted by the
trapped cloud with a photodiode and we stabilise the fluorescence signal to a
reference value by modulating, via the intensity of the 2D-MOT beams, the
flux of cold atoms loading into the 3D-MOT. Under the assumption that the
fluorescence is proportional to the number of atoms, the latter is also
stabilised from run to run.

Once the atoms are loaded in the 3D-MOT, we suddenly switch off the magnetic
field gradient and change the cooler frequency to the value needed for the grey
molasses. Due to technical limitations, this
frequency shift cannot be accomplished instantaneously: we linearly ramp the
cooler frequency in 2 ms. At the beginning of the ramp, we switch off the
(incoherent) repumper light used for the MOT loading and switch on the
(coherent) repumper generated by the EOM sideband for the grey molasses.

As for bright molasses, also for grey molasses the lowest temperatures are reached only after proper
cancellation of stray magnetic fields, and we estimate to reach residual fields below 0.1 G. For this purpose we use three independent pairs of compensation coils placed along the three 3D-MOT axes.

\begin{figure}[h!]
\centering
\includegraphics[width=0.8\linewidth]{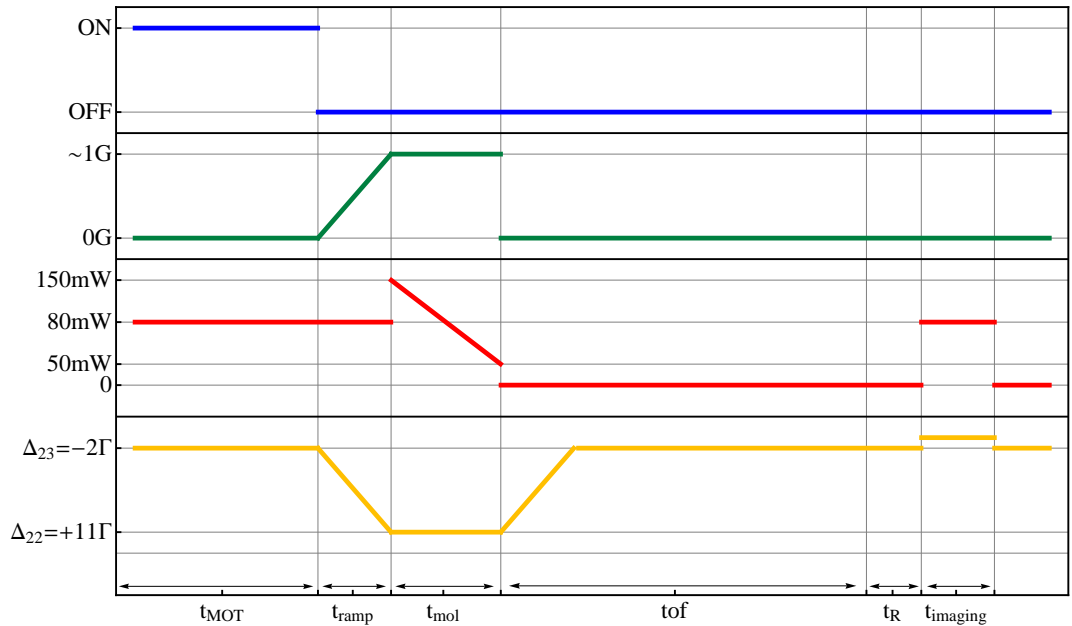}
\captionsetup{justification=justified}
\caption{\textbf{Experimental sequence.} Time dependence of the quadrupole magnetic field (blue), the compensation magnetic field bias (green),  the total optical power of the laser beams (red) and the cooler detuning 
(yellow). We load the MOT for $t_{\text{MOT}}$, then we switch off abruptly the quadrupole magnetic field and ramp the laser frequency and the compensation fields to the molasses value in $t_{\text{ramp}}=2\,\text{ms}$, and perform grey molasses for $t_{\text{mol}}$. Then, we suddenly switch off the lights and let the atoms free to expand for $t_{TOF}$ before acquiring a fluorescence image of the atoms in $F=1$. To do that, before the image we pump the atoms in $F=1$ to the $F=2$ hyperfine state with a repumper pulse of $t_{\text{R}}=200\,\mu\text{s}$, and then switch on the cooler light at the same detuning as the MOT phase - we ramp the frequency back with a second 2 ms ramp when the lights are off during the free expansion - and acquire the image.}
\label{fig:ExpTiming}
\end{figure}

Finally, in \fref{fig:ExpTiming} we report the experimental time sequence, as obtained from the parameters optimization described above. In particular, we have found convenient 
to start the grey molasses with large optical intensity - crucial to capture most part of atoms from the MOT - and then to slowly decrease the beams power with a linear ramp, down to a value lower than the level employed for the MOT loading. 

\subsection*{Fitting procedure: time-of-flight curves}

After the molasses, all the beams are extinguished and the atoms freely
expand. After a few ms of TOF, the MOT beams are switched on
again and the resulting fluorescence emitted from the atomic cloud is collected by the CCD camera. From the recorded images, we measure the size of the cloud $\sigma_i (i=x,z)$  and the number of atoms $N$. 

The dependence of the sizes on the free-fall time duration $t_{\rm TOF}$ is given by the following equation:
\begin{equation}
\centering
\sigma_i\left(t_{\rm TOF} \right)=\sqrt{\left(\sigma_i^0\right)^2+\frac{k_B T_{i}}{m}t_{\rm TOF}^2}
\label{eq:sigmaTOF}
\end{equation}
where the index $i=x,z$ identifies the two spatial directions accessible to fluorescence imaging and $\sigma_i^0$ are the sizes of the cloud immediately after the grey molasses. Therefore, by fitting the fluorescence images of the expanded atomic cloud we determine $\sigma_{i}$ for different values of $t_{TOF}$, and then we extract both $\sigma_i^0$ and $T_{i}$. The temperatures extracted are the same within the error; the temperature $T$ reported in this paper is the average between the values obtained in the two directions.

\subsection*{Data availability}

The datasets generated during and/or analysed during the current study are available from the corresponding author on reasonable request.

\bibliography{sample}

\section*{Acknowledgements}
This work was supported by MIUR (PRIN 2010LLKJBX) and by FP7 Cooperation STREP
Project EQuaM (Grant n. 323714). 
We thank G. Colzi for useful discussions, C. D'Errico for careful reading of the manuscript and  M. Inguscio for enthusiastic support. 

\section*{Author contributions statement}

S.R., S.C., D.N., and F.M. designed and setup the experimental apparatus;  S.R., A.B., C.F., G.R. and F.M. conceived and conducted the measurements; S.R., A.B., C.F. and F.M. analysed the results. F.M. carried out numerical calculations.  All authors reviewed the manuscript. 

\section*{Additional information}

\textbf{Competing financial interests:}  The authors declare no competing financial interests.
 

\end{document}